\documentclass[aps,prd,twocolumn,showpacs,amsmath,amssymb,nofootinbib,eqsecnum, preprintnumbers]{revtex4-1}

\usepackage{color}

\newcommand{\dm}{n}

\newcommand{\be}{\begin{equation}}
\newcommand{\ee}{\end{equation}}
\newcommand{\ba}{\begin{eqnarray}}
\newcommand{\ea}{\end{eqnarray}}
\newcommand{\nn}{\nonumber}

\newcommand{\Mb}{\overline{M}}
\newcommand{\rhob}{\overline{\rho}}

\begin{document}
\title{Generalised Eisenhart lift of the Toda chain}

\author{Marco Cariglia}
\email{marco@iceb.ufop.br}
\affiliation{DEFIS, Universidade Federal de Ouro Preto,  Campus Morro do Cruzeiro, 35400-000 Ouro Preto, MG - Brasil}
\author{Gary Gibbons}
\email{g.w.gibbons@damtp.cam.ac.uk} 
\affiliation{DAMTP, University of Cambridge, Wilberforce Road, Cambridge CB3 0WA, UK}

\date{\today}

\begin{abstract} 
The Toda chain of nearest neighbour interacting particles on a line can be described both in terms of geodesic motion on a manifold with one extra dimension, the Eisenhart lift, or in terms of geodesic motion in a symmetric space with several extra dimensions. We examine the relationship between these two realisations and discover that the symmetric space is a generalised, multi-particle Eisenhart lift of the original problem, that reduces to the standard Eisenhart lift. Such generalised Eisenhart lift acts as an inverse Kaluza-Klein reduction, promoting coupling constants to momenta in higher dimension. In particular, isometries of the generalised lift metric correspond to energy preserving transformations that mix coordinates and coupling constants. A by-product of the analysis is that the lift of the Toda Lax pair can be used to construct higher rank Killing tensors for both the standard and generalised lift metrics. 
\end{abstract}

\pacs{02.30.Ik, 45, 45.10.Na} 
 

\preprint{}

\maketitle
 
\section{Introduction}  
Integrable systems occur rarely in real world applications and yet are extremely valuable theoretically. Apart from the advantage of offering the possibility to study a given system analytically, one of the reasons why integrable systems are important is that they lie at the intersection of a number of research areas, among which dynamical systems, geometrical methods and fundamental theories. 
 
In the study of dynamical systems one of the fundamental questions is to understand what exactly differentiates integrable systems from chaotic ones, especially in the thermodynamical limit. For example it is still an open question to understand what happens to Fermi-Pasta-Ulam systems \cite{fermi1995other_version} for very long time scales: integrable systems like the Toda lattice \cite{Toda1967} and the KdV equation \cite{KdV1895} among others have been used to approximate them. On the other hand, geometry and geometrical methods play an important role, as for example there are several integrable systems in one dimension that can be described in terms of geodesic motion on a symmetric space. Some authors have also conjectured that this might be the case for all one-dimensional integrable systems, for example Perelemov and Olshanetsky \cite{OP1980} and Marmo and collaborators \cite{Marmo1994}. Lastly, integrable systems, even the simplest one dimensional ones, crop up in fundamental theories of physics: notable examples are among others the (super) Calogero model \cite{GaryPaulTownsend1999} that has been argued to provide a microscopic description of the near-horizon limit of the extreme Reissner-Nordstr\"{o}m black-hole, and Toda and Toda-like systems, which appear in the study of gauge field theories \cite{GoddardOliveGanoulis1982,OliveFerreira1985}, supergravity theories \cite{{Chemissany2011}}, and Kaluza-Klein theories  \cite{Gary1982,GaryBreitenlohnerMaison1988}. 
 
In this work we focus our attention on one specific integrable system, the Toda chain. It describes a chain of particles on a 1-dimensional line, interacting via an exponential, nearest neighbour potential. It has been introduced by Toda in \cite{Toda1967}, who showed that it admits solitonic "travelling waves" solutions. Constants of motion where displayed by Henon \cite{Henon1974} and Flaschka \cite{Flaschka1974}. In particular they showed that the Toda chain is a finite dimensional analog of the KdV equation. The Toda chain with $\dm$ particles is one of the special classical Hamiltonian systems described by Olshanetsky and Perelomov in \cite{OP1980} such that their dynamics can be described by geodesic motion on a symmetric space with dimension bigger than $\dm$. For the Toda chain the symmetric space is $X^- = SO(\dm) \backslash SL(\dm, \mathbb{R}) $. At the same time, it is a known result by Eisenhart \cite{Eisenhart1929} that every classical Hamiltonian system with $\dm$ degrees of freedom, kinetic energy of the form $\frac{1}{2}\sum_{i,j=1}^\dm h_{ij}\, \dot{q}^i \dot{q}^j$, in the absence of magnetic field interaction, and with a potential $V$ independent of time, is equivalent to geodesic motion on a space with dimension $\dm+1$ and metric 
\be \label{eq:Eisenhart_metric_generic}
g = \sum_{i,j = 1}^\dm h_{ij} \, dq^i dq^j +  \frac{dy^2}{2V} \, . 
\ee 
In particular for positive potentials, like the one for the Toda chain we are considering in this work, the lift metric is Riemannian. 
The original work by Eisenhart displayed the metric \eqref{eq:Eisenhart_metric_generic} as a special case of a more general construction. When $V$ depends on time, or there is interaction with an external magnetic field, it is possible to consider a Lorentzian lift metric in $\dm +2$ dimensions, and the equations of motion of the original  system are obtained by considering null trajectories. After the original work by  Eisenhart it took a number of years after the same idea was independently re-discovered in \cite{DuvalBurdetKunzlePerrin1985}, from there prompting further work, among which \cite{GaryDuvalHorvathy1991,GaryPeterHorvathyZhang2012}. In recent years the 
technique has been used on several occasions: a non-exhaustive list of examples is given by a geometrical revisiting of integrable motion and its relation to chaos \cite{Pettini2002,PettiniBook}, building of new examples of non-trivial higher order Killing tensors in curved space \cite{GaryDavidClaudeHouri2011,Galajinsky2012}, the geometrical lift of certain types of Dirac equation with flux into a free Dirac equation \cite{Marco2012}. 
 
The relation, if any, between these two types of geometrical lifts of the system was unknown prior to this work. We investigated the problem and found that the Olshanetsky Perelomov lift (OP) in fact corresponds to a generalised Eisenhart lift, where a new dynamical variable $\omega_a$ is introduced for each coupling constant in the system, and $X^-$ is endowed with an $SL(\dm, \mathbb{R})$ right invariant metric. The original Eisenhart lift can be recovered by setting a linear combination of the $\omega$ variables equal to the $y$ variable in \eqref{eq:Eisenhart_metric}, and performing a dimensional reduction to $\dm + 1$ dimensions. A consequence of the construction is that studying the isometries of the metric on $X^-$ corresponds to finding transformations between the coordinates of the original Toda system and the coupling constants such that the energy is left unchanged. Also, as a by-product of the techniques used we have also found new examples of higher rank Killing tensors. 
 
The rest of the work is organised as follows. In sec. \ref{sec:Toda} we present the Toda chain. In sec. \ref{sec:Eisenhart} we discuss the Eisenhart lift of the Toda chain, in particular we obtain new higher rank Killing tensors. In sec. \ref{sec:OP} we present the OP symmetric space lift of the Toda chain, analyse the relevant dynamical variables and find an $SL(\dm, \mathbb{R})$ right invariant metric that induces the correct dynamics. We discuss how a partial dimensional reduction of this metric yields the Eisenhart metric of section \ref{sec:Eisenhart}. We conclude in section \ref{sec:conclusions} with a summary and open questions.

\section{The Toda chain\label{sec:Toda}}  
The non-periodic Toda system with $\dm$ particles is defined by the Hamiltonian 
\be \label{eq:non_periodic}
H(p,q) =  \sum_{i=i}^\dm \frac{p_i^2}{2} + V(q) = \sum_{i=i}^\dm \frac{p_i^2}{2} + \sum_{i=1}^{\dm -1}g_i^2 e^{2 (q_i - q_{i+1})} \, . 
\ee 
It describes a system of particles that interact with their nearest neighbour via an exponential potential. The equations of motion are 
\ba 
\dot{q}_i &=& p_i \, , \nn \\ 
\dot{p}_1 &=& - 2 g_1^2 e^{2(q_1 - q_2)} \, , \nn \\ 
\dot{p}_i &=& - 2 g_i^2 e^{2(q_i - q_{i+1})} + g_{i-1}^2 e^{2(q_{i-1} - q_i)} \, , \;\; i >1 \, . 
\ea

Flaschka in \cite{Flaschka1974} displayed a Lax pair for the system. We are going to use here an alternative version as per \cite{OP1980}, given by the matrices  
\ba 
L_{ij} &=& \delta_{ij} p_j + g_{i-1} \delta_{i,j+1} + g_i \, e^{2(q_i - q_{i+1})} \delta_{i, j-1} \, , \label{eq:Lax_L} \\ 
M_{ij} &=& 2 g_i \, e^{2(q_i - q_{i+1})} \delta_{i, j-1}  \label{eq:Lax_M} \, . 
\ea 
This means that the equations of motion can be rewritten in the form $\dot{L} = [L,M]$. This equation has solution given by  ${L(t)=A(t)L(0)A^{-1}(t)}$, where the evolution matrix ${A(t)}$ is determined by the equation 
\be \label{eq:evolution}
\frac{dA}{dt} = -MA \, . 
\ee 
Therefore, if $I(L)$ is a function of $L$ invariant under conjugation $L\to A LA^{-1}$, then $I(L(t))$ is a constant of motion. In particular one can construct $\dm$ independent conserved quantities $I_i = \frac{1}{2^i} Tr L^i$, $i= 1, \dots, \dm$, where  $I_1 = \sum p_i$, $I_2 = H$. The Toda system is also superintegrable \cite{Damianou2006}. It is important noticing that the properties of integrability and existence of the Lax pair are valid for all choices of the coupling constants. 
 
\section{The Eisenhart lift\label{sec:Eisenhart}} 
As mentioned in the introduction, the Eisenhart lift of the Toda chain is given by an $\dm+1$ dimensional space with coordinates $ q^\mu = \{q_1, \dots, q_\dm, y\}$ and metric 
\be \label{eq:Eisenhart_metric}
g^{(E)} = \sum_{i}^\dm  dq_i^2 +  \frac{dy^2}{2V} \, . 
\ee 
To see the relation to the original Toda chain of section \ref{sec:Toda} one can consider the lifted Lagrangian 
\be 
\mathcal{L} = \frac{1}{2} g^{(E)}_{\mu\nu} \dot{q}^\mu \dot{q}^\nu =  \frac{1}{2} \sum_{i=1}^\dm \dot{q}_i^2 + \frac{\dot{y}^2}{4V} \, , 
\ee 
which yields the momentum $p_y = \frac{\dot{y}}{2V}$, and the Hamiltonian 
\be 
\mathcal{H} = \sum_{i=i}^\dm \frac{p_i^2}{2} + p_y^2 \, V(q) \, . 
\ee 
In particular, since the metric \eqref{eq:Eisenhart_metric} does not depend explicitly on $y$, then $p_y$ is conserved and if we set $p_y = 1$ then we recover the trajectories of the original system. Considering trajectories with $p_y \neq 1$ and non-zero is equivalent to working with a Toda chain with all the coupling constants rescaled by a factor $p_y$. Therefore the Eisenhart lift space corresponds to a collection of Toda chain systems, one related to the other by an overall rescaling of the coupling constants, and the original motion lifts to geodesic motion in dimension higher by one. 
 
Now we can define a lifted Lax pair 
\ba 
\mathcal{L}_{ij} &=& \delta_{ij} p_j + p_y g_{i-1} \delta_{i,j+1} + p_y g_i \, e^{2(q_i - q_{i+1})} \delta_{i, j-1} \, , \nn \\ 
\mathcal{M}_{ij} &=& 2 p_y g_i \, e^{2(q_i - q_{i+1})} \delta_{i, j-1} \nn \, . 
\ea 
Since the original Lax pair was defined for all values of the coupling constants, and since $p_y$ is constant, then it will still be the case that $\dot{\mathcal{L}} = [\mathcal{L},\mathcal{M}]$, and that we can build invariants according to $\mathcal{I}_i = \frac{1}{2^i} Tr \mathcal{L}^i$, $i= 1, \dots, \dm$. However, this time the $\mathcal{I}_i$ are polynomials in the momenta $p_\mu = \{p_1, \dots, p_\dm, p_y\}$ of degree $i$, and therefore they must correspond to rank $i$ Killing tensors $K_{(i)}^{\mu_1 \dots \mu_i}$, $\mu= 1, \dots, \dm+1$, according to the formula 
\be 
\mathcal{I}_i = \frac{1}{i!} \, K_{(i)}^{\mu_1 \dots \mu_i} p_{\mu_1} \dots p_{\mu_i} \, . 
\ee 
These provide to our knowledge new examples of non-trivial higher rank Killing tensors. 
 
Interestingly, the same logic can be applied to create a new lifted Lax pair where each coupling constant $g_i$ lifts to a different momentum $\tilde{p}_i g_i$. Then it is guaranteed one can build conserved quantities for the new Hamiltonian 
\be 
\mathcal{H}_{gen}(p,q) =  \sum_{i=i}^\dm \frac{p_i^2}{2} + \sum_{i=1}^{\dm -1} \tilde{p}_i^2 g_i^2 e^{2 (q_i - q_{i+1})} \, ,  
\ee 
and these quantities will be homogeneous in the momenta. Since the Hamiltonian is quadratic in the momenta it can be interpreted in terms of geodesic motion with respect to an appropriate metric, and therefore we can also construct Killing tensors with respect to the new metric. We call this new metric a \textit{generalised Eisenhart metric}. In the next section we will show that the description of the Toda system given in \cite{OP1980} in terms of geodesic motion in a symmetric space is exactly the motion with respect to the generalised Eisenhart metric.

\section{The Olshanetsky Perelomov lift\label{sec:OP}} 
Olshanetsky and Perelomov \cite{OP1980} showed that the Toda chain equations of motion can also be obtained from geodesic motion on a different space, $X^- = SO(\dm) \backslash SL(\dm, \mathbb{R})$, whose elements are symmetric positive-definite $\dm\times \dm$ matrices $x$ with real components and unit determinant, $\det x = 1$. We broadly follow here the notation they used. 
 
Any real symmetric positive definite matrix $x \in X^-$ admits a Cholesky $UDU$ decomposition, that is it can be written as 
\be \label{eq:n_x_2}
x = Z h^2 Z^T  \, , 
\ee 
where $h^2$ is diagonal and with positive elements, and $Z \in \mathcal{Z} \subset SL(\dm, \mathbb{R})$, the subgroup of upper triangular matrices with units on the diagonal. We can parameterise the diagonal part as 
\be 
h^2 = diag [ e^{2q_1}, \dots, e^{2q_\dm} ] \, , 
\ee 
the $q_i$ being 'dilaton' fields, which will be identified with the positions of the particles in the Toda chain \eqref{eq:non_periodic}. They are restricted by the condition $\det h^2 = 1$, which implies 
\be \label{eq:det1}
\sum_{a=1}^\dm q_a = 0 \, , 
\ee 
which is a condition on the position of the center of mass of the system. 
 
The equation of motion on $X^-$ is given by 
\be \label{eq:geodesic}
\frac{d}{dt} \left( \dot x (t) x^{-1} (t) \right) = 0 \, .  
\ee 
In other words it is an equation for auto-parallel curves with tangent vector $V$ such that $DV /  D t = 0$, where the connection is given by the right connection of $SL(\dm , \mathbb{R})$. In \cite{OP1980} it is shown that if one chooses specific geodesics for which 
\be \label{eq:review_specific_geodesic} 
Z^{-1} \dot{Z} = M \, , 
\ee 
with $M$ given by \eqref{eq:Lax_M}, then \eqref{eq:geodesic} implies the Lax equations $\dot{L} = [L,M]$. 

It is worth noticing that the condition \eqref{eq:review_specific_geodesic}, which selects specific geodesics, is equivalent to say that $Z^{-1}$ satisfies the evolution equation \eqref{eq:evolution}. Then with the current choice of Lax pair we have an evolution equation given by upper triangular matrices. In the work by Flaschka instead a different Lax pair is used and the evolution is given by an orthogonal transformation.

In order to show that the Olshanetsky Perelomov space corresponds to a generalised Eisenhart lift of the Toda chain, that reduces naturally to the Eisenhart lift discussed in sec.\ref{sec:Eisenhart}, we first discuss in the $\dm = 2$ case as a useful example, and then approach the general $\dm$ case.

\section{The $\dm = 2$ case} 
The Lax pair matrices can be written as 
\ba 
\label{eq:2_body_L} L = \left( \begin{array}{cc} p_1 & g_1 e^{2(q_1 - q_2)} \\   g_1 & p_2 
\end{array} \right) \, , 
&& \\ 
\label{eq:2_body_M} M = \left( \begin{array}{cc} 0 &  2 g_1 e^{2(q_1 - q_2)} \\ 0 & 0
\end{array} \right) \, . 
\ea 
We consider the space $X^- = SO(2) \backslash SL(2, \mathbb{R}) $, given by symmetric positive-definite $2\times 2$ matrices with real components 
\be \label{eq:x_generic}
x = \left( \begin{array}{cc} a & c \\ c & b \end{array} \right) 
\ee 
and  unit determinant, $\det x = 1$. If we set $a = \xi_0 + \xi_1$, $b = \xi_0 - \xi_1$ and $c = \xi_2$ then $\det x = 1$ becomes $\xi_0^2 - \xi_1^2 - \xi_2^2 = 1$, which determines the Lobachevsky plane $\mathbb{H}_2$. If $\mathcal{Z}$ is the subgroup in $SL(2, \mathbb{R})$ of upper triangular matrices with units on the diagonal then the matrix $x$ in \eqref{eq:x_generic} can be written as 
\be \label{eq:x_2}
x = Z h^2 Z^T  \, , 
\ee 
with 
\be 
Z =  \left( \begin{array}{cc} 1 & z \\ 0 & 1 \end{array} \right)\, , 
\ee 
and $\det x = \det h^2$. As mentioned above we set 
\be 
h^2 = \left( \begin{array}{cc} e^{2q_1} & 0 \\ 0 & e^{2q_2} \end{array} \right) \, ,  
\ee 
where $\det h^2 = 1$ implies $q_1 + q_2 = 0$. 
 
The auto-parallel equation \eqref{eq:geodesic} is right invariant on $SO(2) \backslash SL(2, \mathbb{R}) $, under the action of the full group $SL(2, \mathbb{R})$. We can build right invariant forms, and from these a right invariant metric. Then, auto-parallel curves will be the same thing as geodesics. To build right invariant forms $\rho^A$ we need to calculate 
\be 
dx \, x^{-1} = \rho^A M_A \, , 
\ee 
where the matrices $M_A$ form a basis for the three-dimensional Lie algebra of $SL(2, \mathbb{R})$ that acts on $X^-$. So we start by anaysing the right invariant term $\dot x (t) x^{-1} (t)$ that appears in the auto-parallel equation \eqref{eq:geodesic}, writing it as 
\ba \label{eq:2_body_intermediate} 
&& \dot x (t) x^{-1} (t) = 2Z \left[ \frac{1}{2} Z^{-1} \dot{Z} + 
\left( \begin{array}{cc} \dot{q}_1 & 0 \\ 0 & \dot{q}_2 \end{array} \right)  \right. \nn \\ 
&& \hspace{-0.7cm} \left. + \frac{1}{2} \left( \begin{array}{cc} e^{2q_1} & 0 \\ 0 & e^{2q_2} \end{array} \right) \left( Z^{-1} \dot{Z} \right)^T \left( \begin{array}{cc} e^{-2q_1} & 0 \\ 0 & e^{-2q_2} \end{array} \right) \right] Z^{-1}  . 
\ea  
From this we find that 
\ba \label{eq:2_body_right_invariant_forms}  \rho^A M_A  &=& 2Z \left[ 
\left( \begin{array}{cc} d q_1 & \frac{1}{2} dz \\ \frac{1}{2} e^{2(q_2 - q_1)} dz & d q_2 \end{array} \right)  \right] Z^{-1}  \nn \\ 
&=& 2Z \left[ 
\left( \begin{array}{cc} \frac{1}{2} d q & \frac{1}{2} dz \\ \frac{1}{2} \frac{g_1^2 dz}{V(q_1)} & - \frac{1}{2} d q \end{array} \right)  \right] Z^{-1}  \, , 
\ea  
where we defined the variables $q = q_1 - q_2$ and  $V(q) = g_1^2 e^{2 q}$. These appear naturally in the reduced Hamiltonian 
\be \label{eq:n2_reduced_Hamiltonian}
H = \frac{1}{2} p^2 + 2g_1^2 e^{2q} \, , 
\ee 
which can be obtained from \eqref{eq:non_periodic} by splitting the original Hamiltonian into a function of $q$ plus a Hamiltonian for the center of mass coordinate $Q = q_1 + q_2$.

We can take the following basis in the Lie algebra of $SL(2, \mathbb{R})$: 
\be 
M_1 = \left( \begin{array}{cc} 0 & 0 \\ 1 & 0 \end{array} \right) \, , 
\ee 
\be 
M_2 = \left( \begin{array}{cc} 0 & 1 \\ 0 & 0 \end{array} \right) \, . 
\ee
\be 
M_3 = \left( \begin{array}{cc} 1 & 0 \\ 0 & - 1 \end{array} \right) \, .  
\ee 
$M_2$ and $M_3$ form a close sub-algebra, $[M_2, M_3] = - M_2$, and the Iwasawa decomposition for $SL(2, \mathbb{R})$ says that for an element of $SO(2) \backslash SL(2, \mathbb{R})$ there is always a representative - different from that displayed in \eqref{eq:x_2} - that can be written as an exponential of $M_3$ times an exponential of $M_2$, so that we can think of $SO(2) \backslash SL(2, \mathbb{R})$ as generated by $M_2$ and $M_3$ through exponentiation. The action of $M_1$ can always be re-absorbed by an $SO(2)$ rotation. 
 
Expanding eq.\eqref{eq:2_body_right_invariant_forms} one gets, modulo constants, the following right-invariant forms: 
\ba 
\rho^1 &=& \frac{g_1^2}{2 V(q)} dz \, ,  \nn \\ 
\rho^2 &=& -  z dq + \left( \frac{1}{2} - \frac{g_1^2 z^2}{2 V(q)} \right) dz \, , \nn \\ 
\rho^3 &=& \frac{dq}{2} + \frac{g_1^2 z}{2 V(q)} dz \, ,   \nn \\ 
\ea 
and the canonically associated right invariant vector fields $R_A$, $A=1,2,3$. The auto-parallel equation \eqref{eq:geodesic} then implies the existence of three constants of motion: 
\ba 
C_1 &=& \frac{g_1^2}{2 V(q)} \dot{z} \, ,  \nn \\ 
C_2 &=& -  z \dot{q} + \left( \frac{1}{2} - \frac{g_1^2 z^2}{2 V(q)} \right) \dot{z} \, , \nn \\ 
C_3 &=& \frac{\dot{q}}{2} + \frac{g_1^2 z}{2 V(q)} \dot{z} \, .  \nn \\ 
\ea  
In particular setting $C_1 = g_1$ we get the condition \eqref{eq:review_specific_geodesic}  with $M$ given by \eqref{eq:2_body_M}. Differentiating $C_3$ one gets the equation of motion for the Toda system upon imposing the condition for $C_1$.  

One can then build the right invariant metric 
\be 
ds^2 = 4 \rho^3 \otimes \rho^3 + 4 \rho^1 \otimes \rho^2 = dq^2 + \frac{g_1^2}{V(q)} dz^2 \, , 
\ee 
and setting $z = \frac{y}{2 g_1}$ one gets exactly the Eisenhart metric 
\be \label{eq:n2_Eisenhart_metric}
ds^2 = dq^2 + \frac{dy^2}{4 V(q)} \, , 
\ee 
that is associated to the reduced Hamiltonian \eqref{eq:n2_reduced_Hamiltonian}. We now consider the general $\dm$ case.

\section{The general $\dm$ case} 
In this section we use the upper and lower triangular matrices $M_{ab}$ and $\overline{M}_{ab}$, and the matrices $M_a$ defined in the appendix.

We can parameterise $Z$ as  
\be 
Z = \exp \left(\sum_{a<b} \omega_{ab} M_{ab}\right) \, . 
\ee 
Since from \cite{OP1980} we expect the relation $\dot{Z} Z^{-1} = M$, the second matrix in the Lax pair, then from the beginning we can make a choice of variables that will induce the condition $\dot{\omega}_{a,a+1} = 2 g_a e^{2(q_a - q_{a+1})}$, $\dot{\omega}_{ab} = 0$ otherwise, and assume that the only non-zero $\omega$s are the $\omega_{a,a+1}$. This amounts to working with a $2(\dm -1)$-dimensional submanifold $\tilde{X}$ of $X^-$, with coordinates $q_a$ and $\omega_{a,a+1}$. We can then write $\omega_a$ instead of $\omega_{a,a+1}$ and write the parameterisation as 
\be 
Z = \exp \left(\sum_{a=1}^{\dm -1} \omega_a M_{a, a+1}\right) \, . 
\ee 
One gets a totally geodesic submanifold of $X^-$ since by construction geodesics on $\tilde{X}$ can be obtained from generic geodesics on $X^{-}$ by imposing $\dot{\omega}_{ab} = 0$ for $b>a+1$. It is convenient to define $\omega_{0} = 0 = \omega_n$. 
 
Using repeatedly \eqref{eq:Mproduct} one finds that for $a<b$ 
\be 
Z_{ab} = \frac{1}{(b-a)!} \omega_a \omega_{a+1} \dots \omega_{b-1} \, 
\ee 
and 
\be 
Z^{-1}_{ab} = (-1)^{b-a} Z_{ab} \, . 
\ee

Now we look for right invariant forms on $\tilde{X}$, invariant under the full group $SL(\dm, \mathbb{R})$, calculating the form $dx x^{-1} = \rho^A M_A$, $A = 1, \dots, \dm^2 -1$, using the generalization of eq.\eqref{eq:2_body_intermediate}. There are three terms, arising from the three terms in square brackets, and we write each one separately. The first one is 
\be 
 Z^{-1} dZ = \sum_{a=1}^{\dm-1} d\omega_a M_{a,a+1} \, , 
\ee 
which is expanded along the $M_{ij}$ matrices defined in section \ref{sec:useful}. These are right invariant forms of the subgroup $\mathcal{Z}$. 
 
The second term from $\dot{x} x^{-1}$ is 
\ba 
&&  2Z diag\left[ dq_1, \dots , dq_{\dm -1},  dq_\dm \right] Z^{-1} \nn \\ 
&& = 2 Z \left( \sum_{a=1}^{\dm-1} dq_a M_a \right) Z^{-1}\, , 
\ea 
where the matrices $M_a$ are defined in the appendix and we have rewritten $dq_\dm$ using \eqref{eq:det1}. Using eq.\eqref{eq:MD} one can write the following expansion for $a<\dm$: 
\be 
Z M_a Z^{-1} = M_a + \sum_{b<c} f_{abc} M_{bc} \, , 
\ee 
and find the coefficients 
\be 
f_{abc} = \delta_{ac} Z_{bc} + (-1)^{c-b} \delta_{ab} Z_{bc} + (-1)^{c-a} Z_{ba} Z_{ac} - \delta_{cn} Z_{bc}  \, . 
\ee 
Then the second term we were looking for is 
\ba 
&& 2 \sum_{a=1}^{\dm-1} dq_a M_a +  2 \sum_{b<c} \sum_{a=1}^{\dm -1} dq_a f_{abc} M_{bc} \nn \\ 
&& = 2 \sum_{a=1}^{\dm-1} dq_a M_a + 2 \sum_{b<c} \Big[ \sum_{a=1}^{\dm-1} d q_a \left( \delta_{ac} Z_{ba} + (-1)^{c-b} \delta_{ab} Z_{ac} \right. \nn \\ 
&& \hspace{2.5cm} \left. + (-1)^{c-a} Z_{ba} Z_{ac} - \delta_{cn} Z_{bc} \right) \Big] M_{bc} \, . 
\ea 
Finally, the third term from $\dot{x} x^{-1}$ is 
\ba 
&& Z  h^2 \left( \sum_{a=1}^{\dm-1} d\omega_a M^T_{a,a+1} \right) h^{-2}  Z^{-1} \nn \\ 
= &&  Z  h^2 \left( \sum_{a=1}^{\dm-1} d\omega_a \Mb_{a+1,a} \right) h^{-2}  Z^{-1} \nn \\ 
= &&   Z   \left( \sum_{a=1}^{\dm-1} e^{-2(q_a - q_{a+1})} d\omega_a \Mb_{a+1,a} \right)  Z^{-1} \, . 
\ea 
We calculate separately 
\ba 
&& Z  \, \Mb_{a+1,a} \,  Z^{-1} \nn \\ 
&& = \Mb_{a+1,a} + \sum_{b<c} g_{abc} M_{bc}  + \sum_{a=1}^{\dm -1} \lambda_{ab} M_b \, , 
\ea 
and find 
\be 
\lambda_{ab} = \delta_{ab} \, \omega_b - \delta_{a+1,b} \, \omega_{b-1} \, , 
\ee 
\ba 
g_{abc} &=&  \delta_{ca} \, Z_{b,a+1} +(-1)^{c-a}  \delta_{ba} \, \omega_a Z_{ac} \nn \\ 
&& + (-1)^{c-a} \delta_{b,a+1} Z_{ac}  \nn \\ 
&& + (-1)^{c-a} \left[  Z_{b, a+1} Z_{ac} - \delta_{ba} Z_{a,a+1} Z_{ac} \right] \, . 
\ea 
Therefore the third term is given by 
\ba 
&&  \sum_{a=1}^{\dm-1} e^{-2(q_a - q_{a+1})} d\omega_a \Mb_{a+1,a} + e^{-2(q_1-q_2)} \omega_1 d \omega_1 M_1 \nn \\ 
&&  + \sum_{a=2}^{\dm-1} \left( e^{-2(q_a - q_{a+1})} \omega_a d\omega_a - e^{-2(q_{a-1} - q_a)} \omega_{a-1} d\omega_{a-1} \right) M_a \nn \\ 
&& + \sum_{b<c} \left(\sum_{a=1}^{\dm -1} e^{-2(q_a-q_{a+1})} g_{abc} \, d\omega_a \right) M_{bc} \, . 
\ea 
We can collect all terms and look for right invariant forms. We write 
\be 
dx x^{-1} = \sum_{a=1}^{\dm-1} \rho_a M_a + \sum_{a<b} \rho_{ab} M_{ab} + \sum_{a>b} \rhob_{ab} \Mb_{ab} \, , 
\ee 
and find first 
\be 
\rhob_{a+1,a} = e^{-2(q_a-q_{a+1})} d\omega_a \, , 
\ee 
the other $\rhob$ forms being zero (on this submanifold of $X^-$). The $\rhob$ forms correspond to the conserved quantities 
\be \label{eq:general_condition_Z_dot_Z_inverse}
\overline{c}_{a+1,a} = e^{-2(q_a-q_{a+1})} \dot{\omega}_a  \, , 
\ee 
and by setting $\overline{c}_{a+1,1} = 2 g_a$ we recover the condition $\dot{Z} Z^{-1} = M$ of \cite{OP1980}, with $M$ defined in \eqref{eq:Lax_M}. Next, there are the forms 
\ba \label{eq:rho_a_forms}
\rho_1 &=& 2 dq_1 + e^{-2(q_1 - q_2)} \omega_1 d\omega_1 \, , \nn \\ 
\rho_a &=& 2 dq_a + \left( e^{-2(q_a - q_{a+1})} \omega_a d\omega_a \right. \nn \\ 
&& \left. - e^{-2(q_{a-1} - q_{a})} \omega_{a-1} d\omega_{a-1} \right) \, , \quad a>1 \, . 
\ea 
These give rise to the conserved quantities 
\ba 
\lambda_1 &=& \dot{q}_1 + g_1 \omega_1  \, , \nn \\ 
\lambda_a &=& 2 \dot{q}_a + g_a \omega_a - g_{a-1} \omega_{a-1}  \, , \quad a>1 \, , 
\ea
which imply the equations of motion arising from \eqref{eq:non_periodic}. The forms $\rho_a$ are defined for $a=1, \dots, \dm -1$. It is however useful to define an $\dm$-th form $\rho_\dm$, which is linearly dependent on the other ones, by setting $a=\dm$ in eq.\eqref{eq:rho_a_forms} above: 
\ba 
\rho_\dm &=&  2 dq_\dm  - e^{-2(q_{\dm-1} - q_{\dm})} \omega_{\dm-1} d\omega_{\dm-1} \nn \\ 
&=& - \sum_{a=1}^{\dm-1} \rho_a \, . 
\ea

Lastly, there are the forms of type $\rho_{ab}$. We don't need to write all of them, the ones we will use are 
\ba \label{eq:forms_rho_a_a_plus_one}
&& \rho_{a,a+1} = 2 \omega_a (dq_{a+1} - dq_a ) + d\omega_a \nn \\ 
&& + \frac{\omega_a}{2} \left[ \left( \omega_{a+1}  e^{-2(q_{a+1} - q_{a+2})} d\omega_{a+1} - \omega_a  e^{-2(q_a - q_{a+1})} d\omega_a \right) \right. \nn \\ 
&& \left. \hspace{-0.5cm} - \left( \omega_a  e^{-2(q_a - q_{a+1})} d\omega_{a-1} - \omega_{a-1}  e^{-2(q_{a-1} - q_a)} d\omega_{a-1} \right) \right]  . 
\ea 
 
Now we build a right invariant metric that will reduce to the Eisenhart lift metric. An explicit calculation shows that 
\ba \label{eq:metric_generic_case} 
g &=& \sum_{a=1}^{\dm} \left(\frac{\rho_a}{2}\right)^2 + \sum_{a=1}^{\dm-1} \frac{\rhob_{a+1,a} \rho_{a,a+1}}{2} \nn \\ 
&=& \sum_{a=1}^{\dm-1 } \left(\frac{\rho_a}{2}\right)^2 + \left( \sum_{a=1}^{\dm-1} \frac{\rho_a}{2} \right)^2 + \sum_{a=1}^{\dm-1} \frac{\rhob_{a+1,a} \rho_{a,a+1}}{2} \nn \\ 
&=& \sum_{a=1}^\dm dq_a^2 + \frac{1}{2} \sum_{a=1}^{\dm-1} e^{-2(q_a - q_{a+1})} d\omega_a^2 \, . 
\ea 
We can define a new variable 
\be 
y= \sum_{a=1}^{\dm-1} g_a \omega_a \, . 
\ee 
On allowed trajectories it satisfies 
\be 
\dot{y} = \sum_{a=1}^{\dm-1} 2 g_a^2 e^{2(q_a-q_{a+1})} = 2 V(q) \, , 
\ee 
where the potential $V$ is given in \eqref{eq:non_periodic}. On the other hand, on trajectories we also have 
\be 
\frac{1}{2} \sum_{a=1}^{\dm-1} e^{-2(q_a - q_{a+1})} \dot{\omega_a}^2 = 2 V(q) \, , 
\ee 
so we have the equality between forms 
\be 
\frac{1}{2} \sum_{a=1}^{\dm-1} e^{-2(q_a - q_{a+1})} d\omega_a^2 = \frac{dy^2}{2V(q)}  
\ee 
on all allowed trajectories. Then this identity holds on the whole span of the trajectories and we can re-write the metric as 
\be 
g = \sum_{a=1}^\dm dq_a^2 + \frac{dy^2}{2V(q)}  \, , 
\ee 
which is the Eisenhart metric. 
 
It is worthwhile noticing that the free particle Hamiltonian associated to the metric \eqref{eq:metric_generic_case} is 
\be 
\mathcal{H} = \sum_{a=1}^\dm  \frac{p_{q_a}^2}{2} + \sum_{a=1}^{\dm-1} p_{w_a}^2 e^{2(q_a - a_{a+1})} \, , 
\ee 
where $p_{w_a} = \frac{1}{2}  e^{-2(q_a - a_{a+1})} \dot{\omega}_a$ is equal to $g_a$ on allowed trajectories, according to eq.\eqref{eq:general_condition_Z_dot_Z_inverse}. This thus corresponds to a multi-particle Eisenhart lift, where for each coupling constant $g_a$ one adds a variable $\omega_a$. Then the isometries of the metric $g$, given by the Killing vectors associated to the right invariant forms of $SL(\dm, \mathbb{R})$, become transformations that leave $\mathcal{H}$ unchanged, and hence are transformations that mix $q$ and $\omega$ coordinates while leaving the energy \eqref{eq:non_periodic} of the original Toda chain unchanged. For example for the conserved quantities explicitly calculated here one has the following cases. First   
\be 
 \overline{c}_{a+1,a} = 2 p_{\omega_a}   
\ee 
correspond to trivial translations of the $\omega$ variables.  
 
Next,  the quantities
\ba 
&& \lambda_1 = p_{q_1} + p_{\omega_1} \omega_1 \, , \nn \\ 
&& \lambda_a = p_{q_a} + p_{\omega_a} \omega_a - p_{\omega_{a-1}} \omega_{a-1} \, , a > 1 \, , 
\ea 
which give transformations of the kind 
\ba 
&& \delta q_1 = \epsilon \, , \nn \\ 
&& \delta \omega_1 =  \epsilon \omega_1 \, , \\ 
&& \delta_{p_{\omega_1}} = - \epsilon p_{{\omega_1}} \, , 
\ea 
and 
\ba 
&& \delta q_a = \epsilon \, , \nn \\ 
&& \delta \omega_a =  \epsilon \omega_a \, , \delta \omega_{a-1} =  - \epsilon \omega_{a-1}  \\ 
&& \delta_{p_{\omega_a}} = - \epsilon p_{\omega_a} \, , \delta_{p_{\omega_{a-1}}} =  \epsilon p_{{\omega_{a-1}}} \, , 
\ea 
where $\epsilon$ is an infinitesimal parameter. 
These correspond to the fact that one can make a constant shift of one of the variables $q_a$ and at the same time a constant rescaling of the $g_a$ and $g_{a-1}$ couplings, while keeping the energy unchanged. Already the conserved quantities associated to the conserved forms of equation \eqref{eq:forms_rho_a_a_plus_one} give rise to more complicated transformations where $\delta p_{\omega_a}$ is not constant. 
 
Summarising, the symmetries of the metric \eqref{eq:metric_generic_case} give all the dynamical transformations between the $q$ and $p_{\omega _a} \sim g_a$ variables that leave the energy unchanged. In particular, the transformations that are of lowest order in the momenta are those given by the Killing vectors, which form an $SL(\dm, \mathbb{R})$ algebra.

\section{Conclusions\label{sec:conclusions}} 
In this work we have examined the relation between the Eisenhart lift of the Toda chain and its higher dimensional description as geodesic motion in a symmetric space. We have found that the symmetric space is a generalised Eisenhart lift of the Toda chain, that dimensionally reduces to the standard Eisenhart lift. In particular, the $SL(\dm, \mathbb{R})$ algebra of isometries of the symmetric space describes transformations that mix the coordinates $q$ with the coupling constants $g$ while keeping the energy constant. We have also constructed new higher rank Killing tensors for the Eisenhart lift metric as a by-product of our analysis. 
 
The generalised Eisenhart lift is a rather general construction that can be performed for any Hamiltonian system that is well defined when its parameters vary in an appropriate open set. In particular, if the system is integrable for all values of the parameters then the generalised lift space should be parallelisable. The concept of transforming coupling constants into dynamical entities resonates in a way with what happens in String/M-Theory, where the expectation values of the dilatons and moduli behave as coupling constants. It would be interesting to know whether this concept can be useful in contexts different from the Toda chain discussed here. 
 
A novel mathematical aspect of this work is that lifting the original Toda Lax pair we have been able to construct higher rank Killing tensors for the lift metrics. This construction seems rather general and can be applied to any integrable system whose Lax pair has homogeneous entries in the momenta and coupling constants, for example the systems described in \cite{OP1980}. Therefore we expect that it can be used to significantly expand the number of known examples of non-trivial higher rank Killing tensors. 
 
An open question that is left unanswered in this work is what is the relation between the Eisenhart lift and the symmetric space lift for the other types of systems discussed in \cite{OP1980}. From the analysis done there it seems that no generalised Eisenhart lift will be involved, since only one coupling constant is appears in the Hamiltonian, but at the same time the dimension of the symmetric space is strictly higher than that of the Eisenhart lift. Therefore, one might wonder whether a correspondence between the two types of lifts still exists, and in positive case whether a completely different mechanism is responsible for it. Answering this question might help clarify the conjecture mentioned in the introduction about the relation between one-dimensional integrable systems and symmetric spaces.

\section*{Appendix: Useful Identities\label{sec:useful}} 
Upper unitriangular matrices $Z$ form a subgroup  of $SL(\dm, \mathbb{R})$, with Lie algebra $\mathcal{Z}$ generated by the strictly upper triangular matrices $M_{ab}$, $a,b = 1, \dots, \dm$, $a < b$, with components 
\be 
\left( M_{ab}\right)_{ij} = \delta_{ia} \delta_{jb} \, . 
\ee 
From the definition it follows that $M_{ab}^2 = 0$. 
The product of two $M$ matrices is given by 
\be \label{eq:Mproduct}
M_{ab} M_{cd} = \delta_{bc} M_{ad} \, . 
\ee 
Given a diagonal matrix $D$, with elements $(D)_{ij} = d_i \delta_{ij}$ (no sum), one gets the product rule 
\be \label{eq:MD}
M_{ab} D = d_b M_{ab} \, , \quad D M_{ab} = d_a M_{ab} \, . 
\ee

We also define the following diagonal, zero trace matrices $M_a$, $a=1, \dots, \dm -1$: 
\be 
M_a = diag \left[ 0, \dots, 1, \dots, 0, -1 \right] \, , 
\ee 
with the $1$ term in the $a$-th position. It is convenient for formulas in the main text to define $M_\dm = 0$. 
 
Similarly we introduce the lower diagonal matrices $\Mb_{ab}$, $a,b = 1, \dots, \dm$, $a > b$, with components 
\be 
\left( \Mb_{ab}\right)_{ij} = \delta_{ia} \delta_{jb} \, . 
\ee 
These also satisfy the product rule 
\be \label{eq:MD}
\Mb_{ab} D = d_b \Mb_{ab} \, , \quad D \Mb_{ab} = d_a \Mb_{ab} \, .  
\ee 
Then we can parameterise the Lie algebra of $SL(\dm , \mathbb{R})$ with the set $\{ M_{ab}, \Mb_{ab}, M_a \}$. 
 
Another useful product rule is 
\be 
M_{ab} \, \Mb_{cd} = \delta_{bc} \left( M_{ad} + \Mb_{ad} + \delta_{ad} \mathbb{I}_a \right) \, , 
\ee 
where $\mathbb{I}_a$ is the diagonal matrix with a $1$ in the $a$-th element and zero otherwise.



\newpage

\newpage

\providecommand{\href}[2]{#2}\begingroup\raggedright\endgroup

\end{document}